\documentclass{appolb}
\usepackage{epsfig,amssymb,amsmath}

\def\bge{\begin{equation}}
\def\ene{\end{equation}}
\def\bg{\begin{eqnarray}}
\def\en{\end{eqnarray}}

\def\e{\epsilon}
\def\bge{\begin{equation}}
\def\ene{\end{equation}}
\def\bg{\begin{eqnarray}}
\def\en{\end{eqnarray}}

\def\e{\epsilon}

\begin{document}
\title{
EMERGENT GAUGE SYMMETRIES IN \\
PARTICLE PHYSICS AND COSMOLOGY
\thanks{Presented at the 65. Jubilee
Cracow School of Theoretical Physics,
Zakopane, June 14-21 2025.}
}
\author{Steven D. Bass 
\address{
Marian Smoluchowski Institute of Physics, 
Jagiellonian University, \\ 
Krakow, Poland \\
\vspace{3ex}
Kitzb\"uhel Centre for Physics, Kitzb\"uhel, Austria}
}
\maketitle
\begin{abstract}
\noindent
Where do gauge symmetries come from? 
These lectures develop the idea that the Standard Model might be emergent, with its gauge symmetries dissolving in some phase transition deep in the ultraviolet. 
The (meta-)stability of the Higgs vacuum may be pointing to some new critical phenomena at very high energy scales, with the Higgs connecting 
physics at LHC laboratory energies to that in the deep ultraviolet. 
In the emergence scenario, the dark energy scale comes out similar to the size of light Majorana neutrino masses. 
These two quantities appear at the same order in a low energy expansion in inverse powers of the scale of emergence, about $10^{16}$ GeV.
Dark matter candidates include axions and phonon like excitations of degrees of freedom above the scale of emergence.
Possible tests of these ideas involve neutrinos as well as gravitational-waves-related signals from the early Universe, which are sensitive to physics at very high energy scales. 
\end{abstract}

\section{Introduction}

Particle physics interactions are governed by the gauge symmetries of the Standard Model (SM)~\cite{Pokorski:1987ed,Taylor:1976ru,Altarelli:2013tya}. 
Where do these gauge symmetries come from?
The SM is working exceptionally well in present particle physics experiments 
from LHC high-energy collider energies through to  
low-energy precision measurements. 
There is no evidence so far in the data for new particles or interactions. 
The SM is a quantum field theory (QFT) built on the gauge groups of U(1)$_{\rm Y} \otimes$ SU(2)$_{\rm L} \otimes$ SU(3)$_{\rm c}$ 
with the gauge bosons being the massless photon of QED and gluons of QCD plus the massive W and Z bosons that mediate the weak interactions.
Within the SM the discovery of the Higgs boson at CERN in 2012 completes the particle spectrum \cite{Bass:2021acr,Jakobs:2023fxh}.
In spite of this success we know that the SM is not the whole story~\cite{Pokorski:2023ukp}.
Open puzzles include the origin of  neutrino masses, 
the matter-antimatter asymmetry in the Universe and issues of dark energy (DE) and dark matter (DM) 
plus primordial inflation at the interface of particle physics with gravitation. 
One also wants to understand the origin of the gauge symmetries and the fermion families, viz. why are there three? 
How high in energy and precision might the SM apply without the need for new interactions and at what energies might the new physics enter?

In looking for deeper understanding, 
two key issues are the 
observed scale hierarchies in particle physics and the stability of the electroweak vacuum.
Naively, one would expect radiative corrections and naturalness arguments to push the Higgs mass to much larger values without extra new physics to stabilise it.
So far, no new physics has been discovered at the LHC at the TeV scale that could work in this direction. 
Quoting the title of an article by Altarelli  \cite{Altarelli:2013lla},
the situation can be described as
{\it ``The Higgs: so simple yet so unnatural.''}
Further, if we assume no coupling to undiscovered new particles, 
the vacuum of the SM sits close to the border of stable and metastable.
This issue is connected to the running of the Higgs self-coupling which can cross zero at a scale deep in the ultraviolet. 
The result is very sensitive to the exact values of SM parameters. 
Small changes in the gauge couplings or the top quark and Higgs boson masses 
can lead to very different physics.
The SM vacuum comes out 
within a few standard deviations of being stable up to the Planck scale hinting at
possible new critical behaviour in the ultraviolet.  
Perhaps the SM is more special than previously believed. 
In seeking to go beyond the present theory, should we be looking for new particles and/or new principles?

Here we explore idea of an emergent SM.
This is the idea that the gauge symmetries of the SM might be ``born" in some phase transition in the ultraviolet \cite{Jegerlehner:2013cta,Jegerlehner:2021vqz,Bass:2021wxv,wspcbook}. 
The long range tail of a statistical system near a critical point behaves as a renormalisable QFT \cite{Wilson:1973jj,Peskin:1995ev}. If the spectrum contains $J=1$ excitations among the quasiparticles, then it is a gauge theory \cite{tHooft:1979hnm}.
Below the phase transition, the physics behaves as an effective theory with characteristic energy equal to the scale of emergence. 
This energy scale will come out about $10^{16}$ GeV in our approach. 
Key phenomenology connects the Majorana neutrino mass scale, with neutrinos being their own antiparticles, and the DE scale.
These terms have similar size and appear at the same order in a low energy expansion \cite{Bass:2020egf,Bass:2020nrg,Bass:2023ece}. 
One also finds interesting ideas for DM in this scenario \cite{Bass:2025tjw}.
Possible tests involve interesting observables with neutrinos and 
with gravitational-waves-related signals from the early Universe.
This article addresses these fundamental issues. More details are given in the book 
\cite{wspcbook}.

Emergence is a fresh paradigm for thinking about the origin of the SM and the open puzzles in particle physics.
Early work in this direction from different starting points 
is discussed in
Refs.~\cite{Bjorken:1963vg,Bjorken:2001pe, 
Jegerlehner:1978nk,
Jegerlehner:1998kt,Jegerlehner:2018zxm,
Forster:1980dg,
Wilczek:1984dh,
tHooft:2007nis}.
Besides in phase transitions, 
emergent gauge symmetries can also arise in renormalisation group decoupling of gauge dependent terms \cite{Wetterich:2016qee} 
and in connection with spontaneously broken Lorentz invariance \cite{Bjorken:2001pe,Chkareuli:2001xe}. 
Recent extra discussion is given in 
Refs.~\cite{Witten:2017hdv,
Verlinde:cern}.
Going further, there are ideas that quantum mechanics and General Relativity might be emergent together.
Any gauge violating terms might decouple
and not propagate in the evolution of quantum fields. That is, unitary evolution might only hold in gauge equivalence classes \cite{tHooft:2007nis}.
The idea that gauge symmetries can be born in the infrared through phase transitions without local order parameters is well known in condensed matter physics, in particular in connection with topological phases of matter and long range quantum entanglement \cite{Powell:2020osu,Zaanen:2011hm,Moessner:book}.
Collective gauge fields beyond the more fundamental
photons of QED can “emerge” from the quantum structure of the many-body ground state. 
Emergent gauge symmetries are important in the 
low-energy limit of the Fermi-Hubbard model 
\cite{Baskaran:1987my,Affleck:1988zz} and ideas about high temperature superconductors \cite{Sachdev:2015slk}, the A-phase of superfluid $^3$He \cite{Volovik:2003fe,Volovik:2008dd}, string-nets \cite{Levin:2004js,Wen:2007joe}, the quantum Hall effect \cite{Tong:2016kpv}
and spin-ice \cite{Rehn:2016eqc}.

What do we mean by emergent symmetry? 
Emergence in physics occurs when a many-body system exhibits collective behaviour in the infrared that is qualitatively different from that of its more primordial constituents as probed in the ultraviolet \cite{Anderson:1972pca,Palacios:book}.
Hadrons like protons, neutrons and pions are emergent from quark-gluon degrees of freedom. 
Chemistry and biology are emergent from electrodynamics.
An interesting case of emergent phenomena from everyday experience is the collective change in the travel direction of starling flocks from individual bird's flight fluctuations.
Symmetries can also be emergent. 
As an everyday example of emergent symmetry, consider a carpet which looks flat and translational
invariant when looked at from a distance. Up close, e.g., as perceived by an ant crawling on it,
the carpet has structure and this translational invariance is lost. The symmetry perceived in the infrared, e.g., by someone looking at it from a distance, ``dissolves'' in the ultraviolet when the
carpet is observed close up.
Gauge symmetries are usually taken as a fundamental input. 
With emergence, 
the gauge principle is governed by collectivity, with large numbers of ``un-gauged" more primordial degrees of
freedom co-operating in unison to generate ``gauged" collective behaviour.

The plan of this article is as follows. 
In Section~\ref{sec:GS} we next give an introduction to gauge symmetries and their role in the SM.
Section~\ref{sec:VS}  describes the important issue of vacuum stability.
Next, Section~\ref{sec:eSM}  explains the idea of an emergent SM.
Connections with the scale hierarchies associated with dark energy and the Higgs mass are discussed in 
Section~\ref{sec:DE}.
Emergent gauge systems in condensed matter physics are reviewed in Sections~\ref{sec:TCM} and \ref{sec:GSCM} with emphasis on details that have parallels with particle physics phenomena.
Finally, in Section~\ref{sec:Conc} we give an outlook and summary of open puzzles in particle physics and cosmology where emergent gauge symmetries might play an essential role and the possible signatures in future experiments.

\section{Gauge symmetries and particle physics}
\label{sec:GS}

The concept of gauge symmetry goes back to Maxwell's theory of electromagnetism.
Its elevation to a dynamical principle leads to prediction of the dynamics of the Standard Model of particle physics.

\subsection{Gauge symmetries and Maxwell's electromagnetism}

Historically, discussions of gauge symmetry started with Maxwell's theory and equations of electromagnetism
\begin{eqnarray}
    \nabla . {\bf B} &=& 0 \nonumber \\
    \nabla \times {\bf E} + \frac{\partial {\bf B}}{\partial t} &=& 0 \nonumber \\
    \nabla . {\bf E} &=& \rho \nonumber \\
    \nabla \times {\bf B} - \frac{\partial {\bf E}}{\partial t} &=& {\bf j}
\end{eqnarray}
Here 
${\bf E}$ and ${\bf B}$ are the electric and magnetic fields; 
$\rho$ and ${\bf j}$ are the charge density and the current. 
In Maxwell's equations
the electromagnetic field can be described in terms of 
vector and scalar potentials
${\bf A}$ and $A_0$ or $\Phi$, viz.  
${\bf E} = -
\partial {\bf A}/
{\partial t} - {\bf \nabla} A_0
$
and 
${\bf B} = \nabla \times {\bf A}$.
These potentials are not unique.
The electric and magnetic fields and hence  
Maxwell's equations which determine the dynamics are invariant under the transformations 
${\bf A} \to {\bf A} + \frac{1}{e}  \nabla \omega$ and
$A_0 \to A_0 + \frac{1}{e}  \frac{\partial \omega}{ \partial t}$ with $\omega(x)$ some arbitrary function
(or in covariant notation 
$
A^{\mu} \to A^{\mu} + \frac{1}{e} \partial^{\mu} \omega
$) 
hinting at some underlying freedom.
This freedom allows one to consider fields subject to various constraints called gauge fixing conditions.
The constraint
$\nabla . {\bf A} =0$ was proposed by Maxwell; 
Lorenz used the form
$\partial^\mu A_\mu = 0$.
With the advent of quantum mechanics, 
Fock observed that 
the wave equations 
are invariant 
under local phase factor transformations 
$\Psi \to \exp({i e \omega/c}) \Psi$
with the addition of the same vector and scalar potentials 
${\bf A}$ and $A_0$ 
that appear with Maxwell's equations.
For the Schrödinger equation this reads
\begin{equation}
    i \hbar \frac{\partial}{\partial t} \Psi ({\bf r}, t) = \biggl[\frac{1}{2m} (-i \hbar \nabla - e {\bf A})^2 + e A_0 \biggr] \Psi({\bf r}, t)
\label{eqn2}
\end{equation}
with these potentials 
making the same gauge transformations that appear with Maxwell's equations.
Weyl elevated this observation to a general principle: gauge invariance!
With the extension to quantum field theories, gauge invariance is taken as the fundamental principle 
driving principle behind the dynamics of elementary particles in Quantum Electrodynamics (QED) and the Standard Model (SM) including electroweak interactions and Quantum Chromodynamics (QCD).
A detailed history of the discovery of gauge theories
is given in Refs.~\cite{ORaifeartaigh:2000dtc,Jackson:2001ia,Straumann:2005hj}.

\subsection{Quantum Electrodynamics}

Quantum Electrodynamics, QED, is described through the Lagrangian
\begin{equation}
    {\cal L} = {\bar \psi} \bigl(i 
    \gamma^{\mu} D_{\mu}
    - m\bigr) \psi - \frac{1}{4} F_{\mu \nu} F^{\mu \nu} .
\label{eqn3}
\end{equation}
Here $\psi$ is the electron field and $A_{\mu}$ denotes the photon.
The gauge covariant derivative 
$
    D_{\mu} \psi = (\partial_{\mu} + ie A_{\mu}) \psi
$
gives
the electron kinetic energy and the electron-photon interaction
with $e$ the electric charge;
$
    F_{\mu \nu} = \partial_{\mu} A_{\nu} - \partial_{\nu} A_{\mu}
$
is the photon field tensor.
The QED Lagrangian can be derived by requiring
invariance under the local U(1) gauge transformation
$
\psi 
\to e^{i \omega(x)} \psi
$
with the kinetic term $\partial_{\mu} \psi$
replaced by the gauge covariant derivative $D_{\mu} \psi$ with the
photon field transforming as
$
A_{\mu} \to A_{\mu} + \frac{1}{e} \partial_{\mu} \omega
$
so that
$ D_{\mu} \psi \to e^{i \omega(x)} D_{\mu} \psi$.
Maxwell's equations are derived from the photon's equations of motion 
\begin{equation}
    \partial_{\mu} 
    F^{\mu \nu} = j^{\nu}
\label{eq:2b}
\end{equation}
with
$j^{\nu} = i e
{\bar \psi} \gamma^{\nu} \psi$.
Gauss's Law here reads as 
${\rm \nabla.E} = \rho$ where 
the electric field
${\bf E} = -
\partial {\bf A}/
{\partial t} - {\bf \nabla} A_0
$
and
$\rho= ie \psi^\dagger \psi$.
The gauge symmetry survives the transition from quantum QED to classical Maxwell electromagnetism.

Quantisation of 
Eq.~(\ref{eqn3}) leads to QED with its Feynman rules and diagrams.
There are some subtleties connecting the photon field $A_\mu$ with physical observables.
First, real photons come with two transverse polarisations whereas  $A_\mu$ has also time and longitudinal components. These need not be considered as physical degrees of freedom.
Some selection of gauge fixing (or constraint on the gauge field) defines the dynamical degrees of freedom in the internal book keeping of practical calculations. One can arrange this keeping only transverse degrees of freedom, though at the expense of manifest Lorentz invariance and gauge covariance in the formalism. 
Also, the photon field is not a Lorentz four-vector. Under Lorentz transformations it picks up an additional gauge term. Gauge invariance of the theory ensures that these terms cancel in observables. The gauge invariant Maxwell's equations are Lorentz covariant. Further, observables such as S-matrix elements are Lorentz covariant and independent of the choice of gauge fixing and gauge degrees of freedom~\cite{Bjorken:1965zz,Weinberg:1995mt}.

\subsection{SU(3) Yang-Mills theory and Quantum Chromodynamics}

The SM theory is built as the mathematical generalisation from QED with 
local U(1) gauge invariance to include 
weak interactions with SU(2) gauge fields acting on left-handed doublets of up and down type 
quarks plus neutrinos and charged leptons
and, also, QCD with SU(3) gluon fields acting on colour quark triplets.
QCD is confining 
with colour singlet hadrons as the external states measured in our experiments. 
Weak interactions are in a Higgs phase with massive W and Z bosons.

QCD is our theory of strong interactions and the structure of hadrons.
It involves the non-abelian Yang-Mills generalisation of 
local gauge invariance from U(1) to SU(3). 
Historically, it developed from the Eightfold Way patterns observed in hadron spectroscopy with wavefunctions described in terms of 
SU(3) flavour, SU(2)  spin and, inside baryons,  
 antisymmetric in a new SU(3) colour label 
plus the parton description of deep inelastic scattering. 
Then came the insight that colour is a dynamical quantum number and the discovery 
of QCD as a non-abelian local gauge theory with coloured gluons as the gauge bosons mediating interactions between quarks and gluons~\cite{Fritzsch:1973pi,Fritzsch:1972jv}. 
In QCD the quark fermions form an SU(3) colour triplet with the theory invariant
under local rotations in SU(3) colour space $\psi \to {\cal U} \psi$,
where ${\cal U}$ is an element of the gauge group.
The gauge covariant derivative is
$
    D_{\mu} \psi = \bigl(\partial_{\mu} 
    + i g_s \frac{\lambda_a}{2}
    A^a_{\mu} \bigr) \psi
$
where $A_{\mu}^a$ are the gluon fields and 
$\lambda_a$ 
are the non-commuting 
Gell-Mann matrices that form the generators of SU(3); 
$g_s$ is the quark-gluon coupling.
Under local SU(3) gauge transformations the gauge covariant derivative transforms as
\begin{eqnarray}
   & & 
   \psi \to {\cal U} \psi
   \nonumber \\
   & & 
   D_{\mu} \to {\cal U} D_{\mu} {\cal U}^{-1}
   \nonumber \\
   & & 
 A_{\mu} (x) \rightarrow A_{\mu} \ ' (x) 
=
{\cal U} A_{\mu} {\cal U}^{-1}  + \frac{i}{g_s} (\partial_{\mu} {\cal U}) {\cal U}^{-1} .
\end{eqnarray}
The gluon field tensor 
$    
G_{\mu \nu} = [D_{\mu}, D_{\nu}]_-
$
transforms as 
$G_{\mu \nu} \to {\cal U} G_{\mu \nu} {\cal U}^{-1}$
and induces non-abelian three- and four-gluon 
interaction
vertices with the gluons carrying colour charge as well as the quarks.
This is a fundamental difference to with QED 
where photons carry no electric charge and 
leads to very different physics.
The three-gluon interaction vertex induces asymptotic freedom \cite{Gross:1973id,Politzer:1973fx} 
whereby the QCD coupling
$\alpha_s(Q^2) = g_s^2/4 \pi$
decreases logarithmically with increasing resolution
or the four-momentum transfer squared 
$Q^2$
that we probe the QCD system with. 
This contrasts with QED where the running value of $\alpha = e^2/4\pi$ 
instead rises logarithmically with increasing $Q^2$. 
(The running couplings for the SM including QCD are shown in Fig.~\ref{fig:runningC} below.) 
Asymptotic freedom implies 
small interaction strength in the ultraviolet and strong interactions in the infrared. 
The QCD coupling is 
commonly believed to saturate at a finite value or renormalisation group (RG) fixed point in the infrared.
Details of quantisation and gauge fixing are given in 
Ref.~\cite{Muta:2010xua}.
QCD is confining.
Only hadrons, colourless bound states of quarks and gluons, exist in the ground state spectrum.
Besides the large infrared coupling, 
non-local gluon topological properties\cite{Shifman:1988zk} may also be important in the confinement process. 
Chiral symmetry becomes dynamically broken with a scalar quark condensate and pions and kaons as the would-be Goldstone bosons.
One finds a confinement radius of about 1 fm.
The glue that binds the proton plays an essential role in its phenomenological properties like its mass and internal spin structure \cite{Bass:2024sgg,Aidala:2012mv,Bass:2004xa,Jaffe:1989jz} plus the rich QCD phase diagram including extensions to finite densities and temperature.

\subsection{The Standard Model and Higgs phenomena}

The SM is built on the gauge groups of
hypercharge U(1) and chiral SU(2) plus QCD SU(3), 
viz. 
${\rm U(1)_Y \otimes SU(2)_L \otimes SU(3)_c}$.
Weak interactions are described by chiral SU(2)
interactions between left-handed quark and lepton doublets.  
mediated by heavy W and Z gauge bosons with range about 0.01 fm. 
The heavy gauge bosons get their masses through the 
Brout-Englert-Higgs (BEH) mechanism
with gauge invariant couplings to a Higgs doublet field built of complex scalar components.  
This Higgs doublet $\Phi$ 
comes with the potential
\begin{equation}
 V(\phi) 
= \frac{1}{2} m^2 \Phi^\dagger \Phi 
+ \frac{1}{4}
\lambda 
(\Phi^\dagger \Phi)^2
\label{eqn6}
\end{equation}
with $m^2 < 0$
signaling classical degenerate minima
and spontaneous symmetry breaking phenomena.
That is, the gauge symmetry of the underlying theory is hidden in the ground state.
The charge neutral weak gauge boson mixes with a U(1) hypercharge gauge 
to make the massless photon and the heavy Z boson. 
The Higgs self coupling 
$\lambda >0$ 
is taken for vacuum stability, 
viz., that the potential should indeed have a minimum.
The Higgs field acquires a vacuum expectation value 
$|\Phi| = v = \sqrt{-m^2/ \lambda} = 
m_{\rm H}/\sqrt{2 \lambda}$ 
at the minima of the potential with $m_{\rm H}$ the Higgs mass. 
Of the four components of the Higgs doublet field,  
three become Goldstone states corresponding to fluctuations around the rim of the potential.
The Standard Model is most transparent when formulated in unitary gauge where the massless Goldstone modes decouple, being  ``eaten" to become the longitudinal modes of the massive W and Z bosons,
conserving the number of degrees of freedom.
The fourth component of the BEH field $\Phi$ is the scalar Higgs particle,
discovered at CERN in 2012 with mass 125 GeV.
Here spontaneous symmetry breaking is defined relative to the choice of gauge, 
e.g. the unitary gauge, 
with all gauge choices being physically equivalent \cite{Kibble:2014gug}.
The BEH mechanism ensures 
renormalisability \cite{tHooft:1971qjg,tHooft:1972tcz,Veltman:1968ki},
with perturbative unitarity 
ensured with the Higgs mass measured at the LHC.
Conversely, consistent high-energy behaviour
requires Yang-Mills structure with massive gauge bosons when one goes beyond massive QED 
\cite{LlewellynSmith:1973yud,Bell:1973ex,Cornwall:1973tb,Cornwall:1974km}.
Renormalisability requires 
gauge anomaly cancellation in the ultraviolet, which groups the fermions into families.
The BEH mechanism with gauge invariant Yukawa 
(Higgs boson to fermion) 
couplings also gives
the masses of the charged fermions in the SM.

The Standard Model couplings and particle masses 
are related.
For the W and Z gauge bosons
\begin{equation}
m_{\rm W}^2 = \frac{1}{4} g^2 v^2 , 
\ \ \ 
m_{\rm Z}^2 = \frac{1}{4} (g^2 + g'^2 )v^2 
\label{eq:2a}
\end{equation}
where 
$g$ and $g'$ are the SU(2) and U(1) electroweak couplings, and 
$v = 246$ GeV is the Higgs vacuum expectation value (vev).
The charged fermion masses are 
\begin{equation}
m_f = y_f \frac{v}{\sqrt{2}}  \ \ \ \ \ 
(f = {\rm quarks \ and \ charged \ leptons})
\label{eq:2b}
\end{equation}
where $y_f$ are the Yukawa couplings.
The Higgs mass is
\begin{equation}
m_{\rm H}^2 = 2 \lambda v^2
\label{eq:2c}
\end{equation}
with $\lambda$ the Higgs self-coupling.
Before neutrino masses, the minimal SM contains 
18 parameters: 
3 gauge couplings 
and 15 in the Higgs sector.  
These SM particle masses range from the electron mass 0.5 MeV to the top quark mass about 173 GeV.
The ultraviolet behaviour of the SM is very dependent on the details of SM parameters. 
The Higgs mass measured at LHC guarantees perturbative unitarity. 
(This was a prime reason for believing the Higgs boson should be found in LHC kinematics.  
Otherwise one would have needed some new dynamics like strongly interacting W bosons \cite{Chanowitz:1998wi,Chanowitz:2004gk}.)
The high-energy behaviour of the SM couplings, especially the Higgs self coupling which is essential for vacuum stability, 
is discussed in Section \ref{sec:VS} below.
At this point, neutrino masses are outside the SM.
There is no right-hand neutrino to make a Dirac type mass term so either right-handed neutrinos are sterile to SM interactions or neutrinos are Majorana fermions meaning they would be their own antiparticles.
Majorana mass terms can be written in terms of the Weinberg operator 
${\cal O}_5 = (\Phi l)^t_i \lambda_{ij} (\Phi l)_j / M + {\bf h.c.}$
where $\lambda_{ij}$ is a flavour mixing matrix and $l$ and $\Phi$ are the lepton and Higgs doublets.
With Majorana neutrinos one finds three possible CP phases instead of just one with three-flavour mixing of Dirac type fermions, e.g., with quarks in the CKM matrix.
\begin{figure}[t!]
    \centering
    \includegraphics[width=0.80\linewidth]{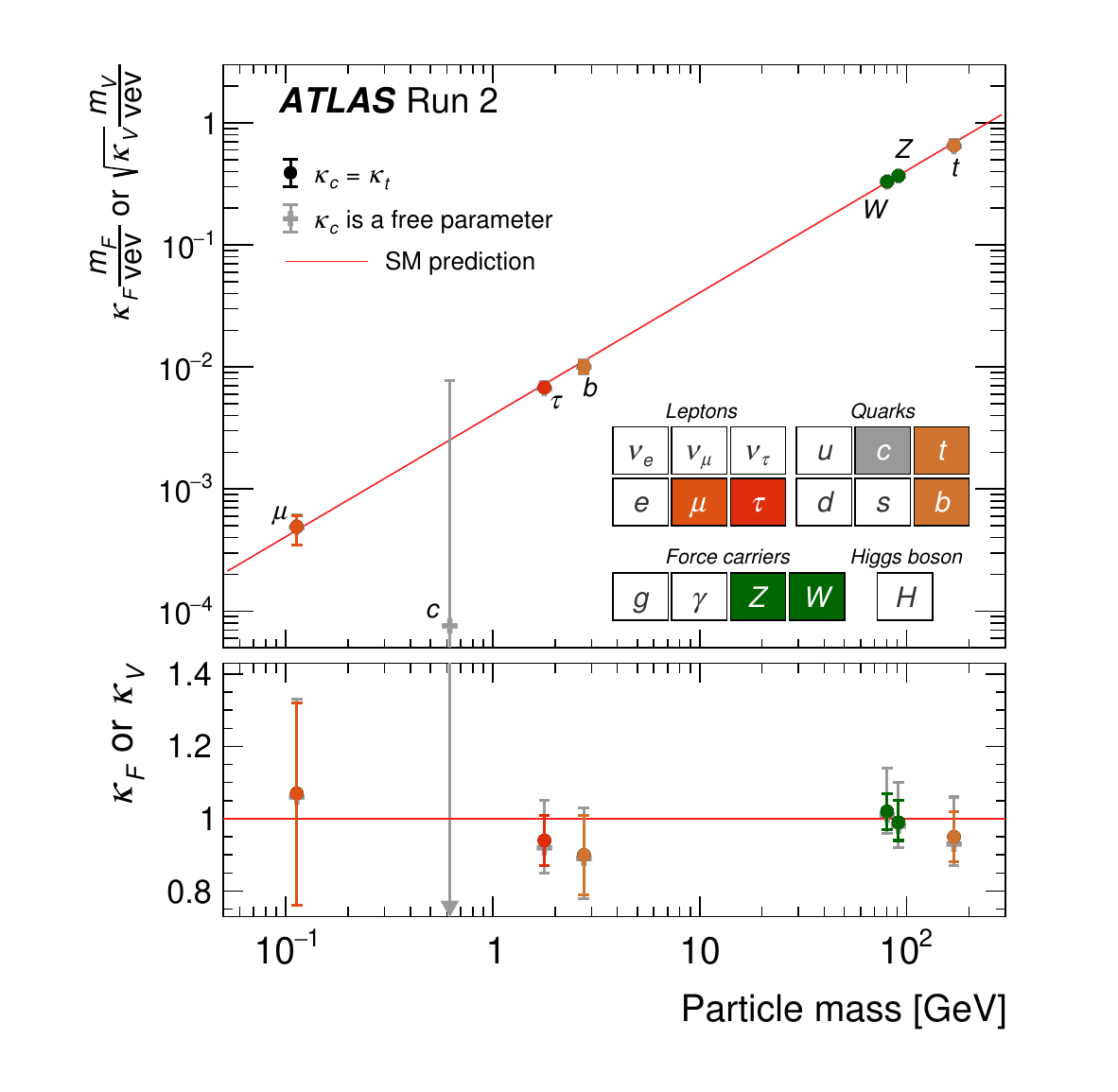}
    \includegraphics[width=0.80\linewidth]{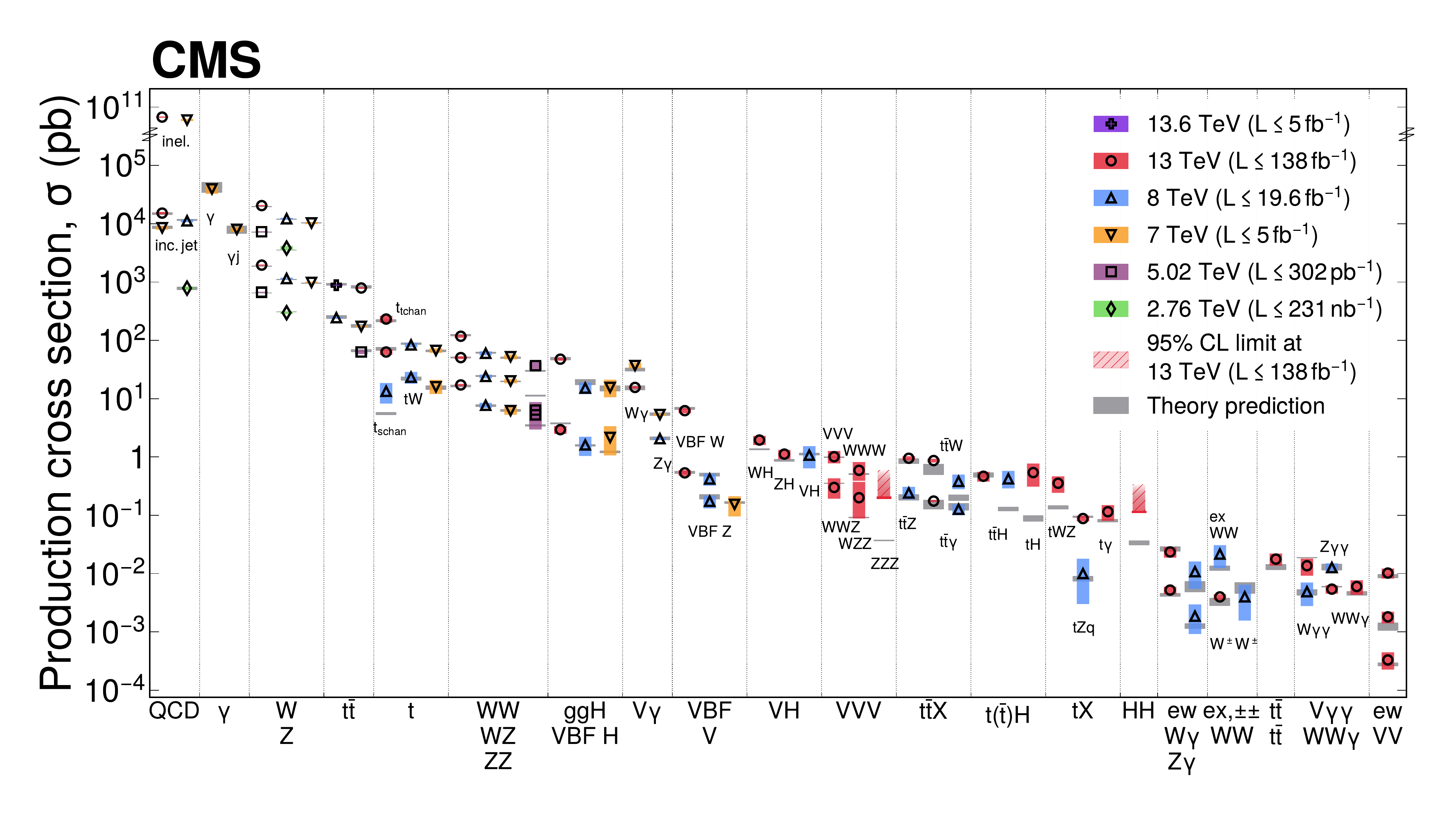}
    \caption{Above: Higgs boson couplings to different particles and masses measured at the LHC. The results are expressed in terms of ``coupling modifiers" which describe the level of deviation from SM expectations \cite{ATLAS:2022vkf}.
Below: Measurements of production cross sections of selected high-energy processes performed at the LHC, all in excellent agreement with the SM theory predictions~\cite{CMS:2024gzs}.}
    \label{fig:SMcouplings}
\end{figure}

There is excellent agreement between SM predictions and the cross sections measured at the LHC.
The Standard Model relations 
Eqs.~(\ref{eq:2a},\ref{eq:2b})
have so far been tested 
at the LHC to about 10\% precision
for the W and Z gauge bosons, the heavy top and bottom quarks, 
and 
the $\tau$ and $\mu$ charged leptons.  
Within present uncertainties, the measured couplings scale as a function of the particle masses just 
as predicted by the Standard Model  
-- see 
\cite{Bass:2021acr,Jakobs:2023fxh}. 
These experimental results are illustrated in  Fig.~\ref{fig:SMcouplings}.
The Higgs self-coupling $\lambda$ awaits accurate measurement.
It is presently known only to the accuracy of a few hundred percent 
~\cite{ATLAS:2025hhd,CMS:2024awa,ATLAS:2024ish} 
(within which it 
 does agree with the SM).
If we take the SM relation 
in Eq.~(\ref{eq:2c}),
$\lambda = m_{\rm h}^2/2 v^2$, then the SM expects 
$\lambda = m_{\rm H}^2/2 v^2 
\approx 0.13$ 
with $v=246$ GeV and $m_{\rm H}=125$ GeV 
at LHC laboratory energies.
(The Higgs vev squared is 
$v^2 = 1/(\sqrt{2} G_F)$ where $G_F$ is Fermi's constant.)
The high luminosity upgrade of the LHC 
should measure $\lambda$ to $\approx 28\%$ precision  \cite{CMS:2025hfp}. 
Going further, 
a 5\% accurate measurement could be made with a future 100 TeV centre-of-mass proton-proton collider~\cite{Jakobs:2023fxh}.
Beyond the Higgs self-coupling, 
one expects an order of magnitude or better improvement in accuracy on the SM parameters from a next generation $e^+ e^-$ collider, 
e.g., the FCC-ee or a future linear collider \cite{deBlas:2025gyz}.
These future experiments will give precision constraints on the SM and hopefully new 
discoveries.
A key issue for understanding deeper physics is the renormalisation group behaviour (RG) of the SM couplings.

\section{Vacuum Stability of the Standard Model}
\label{sec:VS}

In looking for new physics, it is interesting to ask 
how far can we push the SM before needing new particles and/or interactions for consistency reasons. 
In the absence of new physics, it makes sense to extrapolate the SM to the highest scales and to look for consistency issues.
The theoretical extrapolation is performed using RG evolution.
If we assume no coupling to undiscovered new particles, 
then the SM remains finite and well behaved with no Landau pole singularities below the Planck scale.
That is, the Standard Model is mathematically consistent up to the Planck scale.
Further, 
with no couplings to extra particles at higher energies,
the Standard Model revealed by current experiments becomes strongly correlated with its behaviour in the extreme ultraviolet through vacuum stability considerations.
This may be telling us 
something deeper about
the origin of the Standard Model.

LHC data, while so far not revealing any evidence for new particles or interactions, does come with the fascinating issue of vacuum stability.
SM Higgs vacuum stability means that the Higgs self-coupling is positive. 
But RG dependence tells us that  
the Higgs self-coupling decreases with increasing energy scale.
The SM gauge couplings as well as the Higgs 
self-coupling and fermion Yukawa couplings (plus accompanying particle masses and the Higgs vev) 
are scale dependent under RG evolution.
In the pure SM one finds that 
the Higgs self-coupling $\lambda$ stays positive under renormalisation group evolution up to at least $10^{10}$ GeV and maybe up to the Planck scale if we assume just the SM with its measured masses and couplings and no coupling to new particles or interactions \cite{Jegerlehner:2013cta,Bednyakov:2015sca,Degrassi:2012ry,Buttazzo:2013uya,Alekhin:2012py,Masina:2012tz,Hiller:2024zjp}.
With the SM parameters measured at LHC, the SM vacuum is within 
a few standard deviations 
of being stable up to the Planck scale~\cite{Bednyakov:2015sca} 
with the vacuum close to the border of stable and metastable.
The vacuum is stable if the Higgs self-coupling is positive definite up to the scale of ultraviolet completion, 
which we take either as the characteristic energy of the SM if it should be treated as an effective theory or as the Planck mass $M_P = 1.2 \times 10^{19}$ GeV if it can be continued to the maximum possible scale where quantum gravity might become important.
The vacuum becomes metastable if $\lambda$ crosses zero 
with a new minimum in the effective Higgs potential not far below the scale of ultraviolet completion.
Whether this 
happens is very sensitive to exact values of SM parameters, especially the top quark and Higgs masses and to details of higher order radiative corrections.

\begin{figure}[t!]  
\centerline
{\includegraphics[width=1.0\textwidth]
{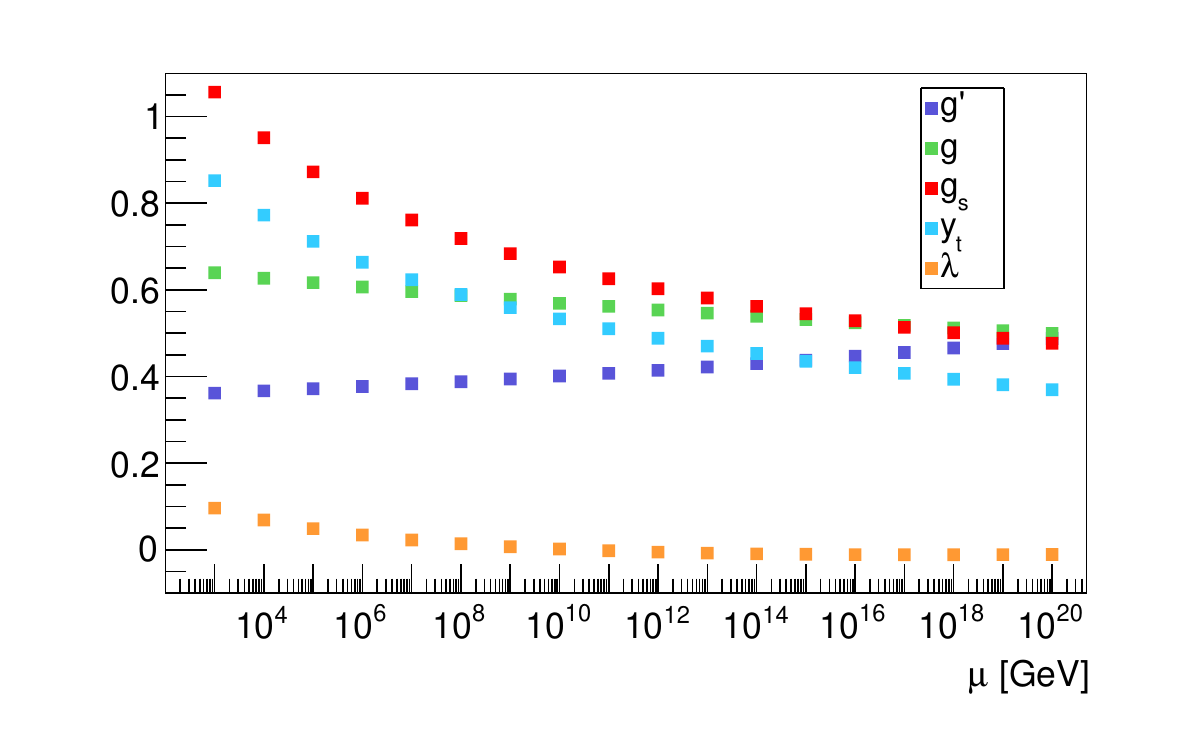}
}
\caption{Running of the Standard Model
gauge couplings $g$, $g'$, $g_s$
for the electroweak SU(2) and U(1) and colour SU(3), 
the top quark Yukawa coupling $y_t$ 
and Higgs self-coupling $\lambda$.
(From left, the points describe the evolution of
$g_s$, $y_t$, $g$, $g'$, $\lambda$ in descending order.)
Figure taken from \cite{Bass:2020nrg}.}
\label{fig:runningC}
\end{figure}

RG evolution calculations take as inputs the measured values of the SM couplings and masses at LHC energy scales.
One also needs input on the Higgs self-coupling $\lambda$. 
In the absence of direct measurement, 
one assumes the Standard Model relation connecting 
$\lambda$ to the Higgs mass, 
Eq.~(\ref{eq:2c}) 
with
$\lambda = m_{\rm H}^2/2 v^2 
\approx 0.13$ 
as input at LHC laboratory scales. 
Calculations have been  performed with the SM 
evolved up to the Planck scale with the measured masses and couplings as input and using three-loop
RG, two-loop matching plus pure QCD corrections evaluated to four loops. 
Results for the SM 
running couplings performed with the 
RG evolution package 
\cite{Kniehl:2016enc}
are shown in 
Fig.~\ref{fig:runningC}. 
The QCD and electroweak SU(2) 
couplings are asymptotically free, decaying 
logarithmically 
with increasing resolution,  
whereas
the U(1) coupling is non-asymptotically free rising in the ultraviolet. 
These couplings almost meet in the ultraviolet but don't quite. 
The top quark Yukawa coupling decreases with increasing resolution.
In general, a higher 
top quark mass tends to reduce $\lambda$ deep in the ultraviolet whereas a larger Higgs mass tends to increase it. 
Sensitivity to QCD corrections involving 
$\alpha_s$ 
means sensitivity also 
to the numbers of colours and active flavours and the QCD scale $\Lambda_{\rm QCD}$. 
Both electroweak and QCD physics thus enter the vacuum stability calculations.
One finds that $\lambda$ stays positive up to at least about $10^{10}$ GeV 
for a top mass $m_t = 173$ GeV.
The exact details where $\lambda$ might cross zero are calculation dependent. 
Within the calculations of Ref.~\cite{Bass:2020nrg}, $\lambda$ would stay positive up to the Planck scale with a top mass $m_t = 171$ GeV with the Higgs mass kept fixed at 125 GeV.
With the measured top quark mass $m_t$
and QCD coupling $\alpha_s$,  
the Standard Model 
needs a Higgs mass $m_{\rm H}$ 
bigger than about 125 GeV 
to ensure vacuum stability, 
making the Higgs particle discovered at CERN especially interesting.
(The Higgs self-coupling would generate a Landau pole singularity if the Higgs mass were about 30\% larger with the measured value of $m_t$, placing a theoretical perturbative upper bound on 
the possible Higgs mass~\cite{Hambye:1996wb}.)
The most metastable solution with $\lambda$ crossing zero around $10^{10}$ GeV corresponds to a life time $\sim 10^{600}$ years~\cite{Buttazzo:2013uya}, much greater than the present age of the Universe.

Modulo the large renormalisation group extrapolation involved, 
the SM vacuum sitting close to the border of stable and metastable might be hinting at some possible new critical phenomena in the ultraviolet \cite{Degrassi:2012ry,Jegerlehner:2013cta,Buttazzo:2013uya}. 
There are ideas that 
near-criticality 
might act as an attractor point in evolution of the higher energy phase, 
with analogous systems in Nature discussed in ~\cite{Buttazzo:2013uya}.
With just small changes in the Standard Model parameters the emerging low energy theory would be very different from the Standard Model.
Vacuum stability involves a 
delicate fine tuning and conspiracy of SM parameters.
The mass of the Higgs boson might be environmentally determined, linked to the stability of the electroweak vacuum. 
If SM vacuum stability is taken as a guiding principle, then the  
SM parameters may be correlated with physics deep in the ultraviolet with an implicit reduction in the number of fundamental couplings.
The SM when taken alone then comes with three key scales: the QCD and electroweak scales, 
$\Lambda_{\rm QCD} \approx 300$ MeV and
$\Lambda_{\rm ew} \approx 246$ GeV respectively, 
plus a large scale in the ultraviolet
where the Higgs self-coupling crosses zero 
(assuming it does so below the Planck scale)
which is linked to the stability of the  vacuum and which might be taken as the scale of ultraviolet completion if we require the vacuum to be fully stable.

Going further, as a small numerical correction, if we consider the SM as an effective theory valid up to the characteristic energy $M$, 
then the usual SM will be supplemented 
by a tower of higher dimensional operator terms suppressed by powers of $1/M$. 
These terms become important only when we approach energy  scales close to $M$. 
They may induce changes
in the behaviour of $\lambda$ close to these energies and thus the exact value where $\lambda$ might cross zero; 
for extended discussion see Ref.~\cite{Branchina:2013jra}.
In models that go beyond the SM with an extended Higgs sector, the Higgs vacuum can exhibit a more complicated structure with the more 
SM-like Higgs having very different self-coupling values, perhaps triggering a first order phase transition in the early Universe~\cite{Biekotter:2022kgf}. These models will be probed in future collider experiments.

\section{An emergent Standard Model}
\label{sec:eSM}

Usually, the gauge symmetries of the SM are put in by hand.
Gauge symmetries 
can be emergent below some large scale in the ultraviolet.
There are ideas connected to phase transitions, to renormalisation group decoupling of gauge symmetry violating terms and to spontaneously broken Lorentz symmetry.

A statistical 
system near its critical point has the interesting feature that its
long range asymptote behaves as a quantum field theory with properties described by the renormalisation group~\cite{Wilson:1973jj,Peskin:1995ev}.
Here, the famous example connects Ising like systems and $\phi^4$ theory with the critical dimension equal to four.
In more than four dimensions the long range asymptote is equivalent to a free field theory.
If the low-energy phase includes $J=1$ excitations among 
the degrees of freedom, then it is a gauge theory. 
(Renormalisable quantum field theories with $J=1$ vector particles 
exhibit local gauge symmetries
\cite{tHooft:1979hnm}.) 
In this case the 
gauge symmetries would be an emergent property of the 
low-energy phase and ``dissolve'' in a phase transition deep in the ultraviolet 
~\cite{Jegerlehner:2013cta,Bass:2021wxv,Bass:2023ece,Bjorken:2001pe,Jegerlehner:1978nk,Jegerlehner:1998kt,Forster:1980dg,tHooft:2007nis}.
If the SM might work this way, 
the quarks and leptons as well as the gauge bosons and
Higgs boson would be the stable collective long-range excitations of some (unknown) more primordial degrees of freedom that exist above the scale of emergence.
Small gauge multiplets would be preferred as the easiest to form as collective excitations \cite{Jegerlehner:2013cta}.

An emergent SM would 
behave as an effective theory.
The long range interactions would be described 
by the 
SM Lagrangian 
which gives a 
renormalisable theory up to  
mass dimension $D=4$.
The minimal SM would be 
supplemented 
by a tower of non-renormalisable higher dimensional operators, 
each 
suppressed by powers of the large scale of emergence~\cite{Jegerlehner:2013cta,Witten:2017hdv}. 
At $D=4$ 
the global symmetries 
are constrained by gauge invariance and renormalisability.
The higher dimensional operators are less constrained and may exhibit extra global symmetry breaking. 
These terms become important 
at energy scales close to the scale of emergence.
At low energies their effect is suppressed by powers of the cut-off.
If we want a stable ground state below the phase transition, 
then the Higgs vacuum considerations discussed in Section \ref{sec:VS} imply constraints on the possible values of the SM parameters.
The Higgs mass might be environmentally selected in connection with the stability of the vacuum.

Lepton number violation and tiny Majorana neutrino masses may enter at $D=5$, 
viz. 
suppressed by a single power of
the scale of emergence, through the so-called Weinberg operator 
\cite{Weinberg:1979sa}. 
This yields neutrino masses 
\begin{equation}
    m_\nu \sim \Lambda_{\rm ew}^2/M
\label{eq:numass}
\end{equation}
where 
$\Lambda_{\rm ew}$ 
is the electroweak scale and $M$ is the scale of emergence. 
Neutrinos would be Majorana fermions meaning that they would be their own antiparticles.
Majorana neutrinos 
and the accompanying lepton number violation could be looked for in neutrinoless double $\beta-$decay experiments. 
With Majorana neutrinos there are 3 extra CP phases that can enter.
New CP violation, needed for baryogenesis, might occur in 
these Majorana phases at $D=5$ and also in new $D=6$ operators 
\cite{Grzadkowski:2010es}, 
which are suppressed by two powers of $M$.
Proton decays~\cite{Weinberg:1979sa,Wilczek:1979hc}.
and Lorentz symmetry violations~\cite{Bjorken:2001pe} might also occur at $D=6$.
As a possible modification of QED, 
the $D=5$ Pauli term 
$\frac{e}{M} {\bar \psi} \sigma_{\mu \nu} \psi F^{\mu \nu}$  
would give an extra contribution to the electron's anomalous magnetic moment. 
Light mass pseudoscalar axion particles have been proposed in connection the strong CP puzzle~\cite{Weinberg:1977ma,Wilczek:1977pj}.
If present, 
these come with masses and couplings to SM particles entering at $D=5$. 
The fact that global symmetries like lepton and baryon number conservation 
plus Lorentz invariance are working so well in experiments tells us that any violation should arise deep in the ultraviolet.
Extra discussion of the SM as an effective theory including higher dimensional operators is given in Ref.~\cite{Weinberg:2018apv}.

Phenomenological constraints 
on the scale of emergence in this scenario 
come from neutrino masses 
(and also dark energy, see Section \ref{sec:DE}). 
Neutrino oscillation experiments, where neutrinos created with a particular flavour are later measured to have a different flavour, 
point to the existence of tiny neutrino masses.
Assuming three species of neutrinos, the neutrino oscillation data 
constrains the largest mass squared difference to be
$\approx 2 \times 10^{-3}$ eV$^2$ 
with the smaller one as 
\hbox{$(7.53 \pm 0.18) \times 10^{-5}$ eV$^2$
\ 
\cite{BahaBalantekin:2018ppj}}.
With these values the lightest neutrino mass is expected to be about 10$^{-8}$ times the value of the electron mass.
While suggestive, 
neutrino oscillation experiments measure only neutrino mass differences.
They do not tell us the absolute mass scale or whether neutrinos are
Majorana or Dirac.
Direct probes of absolute neutrino masses are 
neutrinoless double $\beta$-decay experiments and the KATRIN experiment which 
provides a clean, model-independent anchor for the light neutrino mass scale. 
KATRIN studies the endpoint of the tritium $\beta$-decay
spectrum. 
They measure an effective electron-neutrino mass \cite{KATRIN:2024cdt}  with the experiment aiming at sensitivity to 
$m_\nu \sim 0.3$ eV or less. 
Next generation 
neutrinoless double $\beta-$decay experiments should be sensitive to a Majorana mass parameter down to about 0.01 eV \cite{Baudis:2023pzu}.
If neutrino masses are described via the Weinberg operator or see-saw mechanism, then observations at these levels
together with the constraints from neutrino oscillation experiments would imply very high mass scales.
Otherwise, some 
alternative mass generation mechanisms would be required. 
If we take 
$m_\nu \sim 10^{-8} m_e$, 
for the lightest neutrino mass,
then 
this neutrino mass corresponds to 
$M \sim 10^{16}$ GeV 
when one substitutes in Eq.~(\ref{eq:numass}).
It is interesting that a scale of emergence 
$M \sim 10^{16}$ GeV is within the range where $\lambda$ might cross zero in the vacuum stability calculations discussed in Section~\ref{sec:VS}.
LHC data (so far) reveal no evidence for higher dimensional correlations in searches for new operator terms divided by powers of a large mass scale below the few TeV range \cite{Slade:2019bjo,Ellis:2020unq}.
Present electron anomalous magnetic moment measurements \cite{Fan:2022eto}
suggest a constraint 
$M > 4 \times 10^9$ GeV to avoid a new Pauli term contribution in the domain where the experiments are in agreement with QED at $D=4$.
Increased precision with neutrino and electron magnetic moment experiments will come with development of new quantum sensing technologies \cite{Bass:2023hoi}.

Within the picture of an  emergent SM, if one can increase the energy much above the electroweak scale then the physics becomes increasingly  symmetric with energies 
$E \gg \Lambda_{\rm ew}$ 
until we come within about 0.1\% or so of the scale of emergence. 
At this energy, 
new global symmetry violations from the higher dimensional operators would become important so the physics becomes increasingly chaotic as we approach the phase transition associated with the scale of emergence.
The physics above this scale 
would be described by new degrees of freedom and perhaps new physical laws.
This means that any perturbative extrapolation of SM degrees of freedom above the scale of emergence would reach into an unphysical region.
This scenario contrasts with unification models which involve maximum symmetry in the extreme ultraviolet. 
In unification models the gauge couplings of the SM would meet at some large scale. 
They do almost but not quite -- see Fig.~\ref{fig:runningC}.
The tower of higher dimensional operators here that becomes increasingly more active as we approach the energy scale of the phase transition has a parallel in QCD. There, a tower of higher twist operators involving 
quark-gluon correlations becomes increasingly more important at lower momentum transfers in, e.g., deep inelastic scattering as one gets closer to the confinement transition between quark/gluon and hadronic degrees of freedom.

We have highlighted the idea of an emergent SM ``born'' in some phase transition at very high scales.
Emergent gauge symmetries can also appear through the decoupling of gauge violating terms 
in the RG at an infrared fixed point 
\cite{Wetterich:2016qee}
and also in connection with possible spontaneous breaking of Lorentz symmetry (SBLS)~\cite{Bjorken:1963vg,
Bjorken:2001pe,Chkareuli:2001xe}.
In the former case, 
the coefficient of any local gauge symmetry violating term blows up at the fixed point, 
in contrast to restoration of global symmetries where the coefficient of any symmetry violating term goes to zero at the fixed point. 
With SBLS, 
non observability of any Lorentz violating terms at $D=4$ 
corresponds to gauge symmetries for vector fields like the photon. 
Possible Lorentz violation here might be manifest in terms at largest with size ${\cal O} (\Lambda_{\rm ew}^2/M^2)$ 
with a preferred reference frame naturally identified with 
the frame where the cosmic microwave background, CMB, 
is locally at rest \cite{Bjorken:2001pe}.

There are also ideas that quantum mechanics might be emergent along with the local gauge symmetries of particle physics as well as gravitation \cite{tHooft:2007nis}.
Quantum operators might emerge from the long distance behaviour of a statistical system, perhaps cellular automata 
\cite{tHooft:2010mdz,Wetterich:2009tr,Wetterich:2021exk,Wetterich:2021san}
with symmetry properties not present in the more primordial system. 
In this case, unitary evolution might only hold in gauge equivalence classes (including the equivalence classes generated by general coordinate transformations) so that local gauge invariance becomes an emergent symmetry. Information about the gauge parameters would not propagate in time and decouple in the evolution of quantum fields.
If General Relativity is emergent at a scale below the Planck mass, this would eliminate the well known problems with quantising gravitation.
Quantum theory is believed to work up to scales at least $10^{16}$ GeV \cite{Calmet:2015lsa}.

\section{Scale hierarchies: dark energy and the Higgs mass}
\label{sec:DE}

Particle physics and cosmology comes with the following hierarchies of scales
\begin{equation}
    \mu_{\rm vac} \ll \Lambda_{\rm ew}, m_{\rm H} \ll M_{\rm Pl} .
    \label{eq5}
\end{equation} 
Here, $\Lambda_{\rm ew} \approx 246$ GeV is the electroweak scale, the dark energy scale is 
$\mu_{\rm vac} = 0.002$ eV (see below) and 
$M_{\rm Pl} = 1.2 \times 10^{19}$ GeV is the Planck scale where quantum gravity effects might become important. 
The Higgs mass and the zero-point energies (ZPEs) of quantum field theory (which in principle contribute to the cosmological constant) come with 
quadratic and quartic divergences respectively when radiative corrections evaluated in four dimensions in momentum space.
If taken alone, these contributions would give very large corrections involving powers of the scale of ultraviolet completion for the SM effective theory, perhaps as high as the Planck mass.
Further, without some extra symmetry property like gauge invariance (with spin-one gauge bosons) and chiral symmetry (with fermions),  effective theory arguments would push the energy 
scale of scalar observables towards the characteristic energy for the effective theory, that is, deep in the ultraviolet.
Hence one has the question: Why are the measured values ``so small" compared to the maximum possible energy scales one might consider?
Emergence ideas can help explain these scale hierarchies in terms of 
the space-time translational invariance of the vacuum 
for the dark energy scale 
and vacuum stability for the Higgs mass.

\subsection{Dark energy and an emergent SM}

The accelerating expansion of the Universe is 
driven by dark energy. 
The simplest explanation is a
cosmological constant in Einstein's equations of General Relativity 
connected to the vacuum energy perceived by gravitation~\cite{Weinberg:1988cp,Peebles:2002gy}. 
Einstein's equations 
read as
\begin{equation}
R_{\mu \nu} - \frac{1}{2} g_{\mu \nu} \ R = 
- \frac{8 \pi G}{c^4} T_{\mu \nu} + \Lambda g_{\mu \nu} .
\label{eq2}
\end{equation}
Here $R_{\mu \nu}$ is the Ricci tensor, 
$R$ is the Ricci scalar
and 
$T_{\mu \nu}$ is the energy-momentum tensor 
for excitations above the vacuum;
$G$ is Newton's constant and $c$ is the speed of light. 
The cosmological constant 
term $\Lambda$ 
enters proportional to the metric tensor $g_{\mu \nu}$
and may be interpreted  terms of the vacuum energy density perceived by gravitation 
\begin{equation}
\rho_{\rm vac} = \Lambda 
\times 
 c^4 
/ (8 \pi G ) 
\label{eq3}
\end{equation}
with an associated scale $\mu_{\rm vac}$,
$\rho_{\rm vac} = \mu_{\rm vac}^4$.
The cosmological constant in General Relativity comes with the vacuum equation of state (EoS) 
{\it energy density = - pressure}.
A positive cosmological constant gives accelerating expansion of the Universe (before possible time dependence of dark energy).
Astrophysics observations \cite{Planck:2018vyg} 
tell us 
that $\Lambda = 1.088 \times 10^{-56} \ {\rm cm}^{-2}$ corresponding to a cosmological constant scale 
(in natural units)
\begin{equation}
\mu_{\rm vac} = 0.002 {\rm \ eV} .
\label{eq4}
\end{equation}
The present period of accelerating expansion  
began about five billion years ago when the matter density of the expanding Universe fell below 
$\rho_{\rm vac}$, which then took over as the main  driving term for the expansion.
The Universe has a  spatially flat geometry today on large distance scales.

The cosmological constant  scale in 
Eq.~(\ref{eq4}) is very small compared to usual particle physics scales that characterise the 
SM particle physics vacuum.
A priori, 
one expects the vacuum energy to be sensitive to quantum fluctuations 
through ZPEs  
and potentials in the vacuum with these terms involving the much larger QCD and electroweak scales.
One also has an extra ``bare gravitational contribution'' 
$\rho_\Lambda$ 
to the net $\rho_{\rm vac}$ which then becomes a sum over ZPE, potential and bare gravitational terms, viz.
\begin{equation}
    \rho_{\rm vac} = \rho_{\rm zpe} + \rho_{\rm potential} + \rho_{\Lambda} .
\label{eqn15}
\end{equation}
The net cosmological constant as an observable is independent of any particle physics renormalisation scale whereas 
each of the three terms in Eq.~(\ref{eqn15}) is separately dependent on the renormalisation scale.
Whereas the net cosmological constant comes with a vacuum EoS, 
the EoS satisfied by the individual terms is 
dependent on the symmetry details of the ultraviolet regularisation used in calculating them.
For example, the ZPE can obey a non-vacuum EoS if one chooses a non-covariant regularisation, e.g., a brute force cut-off on momentum integrals, which must be compensated in the bare gravitational term to ensure that the net $\rho_{\rm vac}$ obeys the correct vacuum EoS.  
Detailed discussion is given in Ref.~\cite{Bass:2023ece}.

In the context of the SM and General Relativity 
vacuum energy only becomes an observable when coupling to gravitation.
Usual particle physics  related processes measure differences between quantities rather than absolutes. 
This allows, e.g., normal ordering to set the particle physics vacuum energy to zero 
(before considerations of spontaneous symmetry breaking).
The Casimir effect sometimes taken as evidence of ZPEs but can be calculated without recourse to them \cite{Jaffe:2005vp}.
Within the SM coupled to GR, vacuum energy appears as a cosmological constant  term in Einstein's equations where it drives the accelerating expansion of the Universe.
If the vacuum carries energy, then it gravitates.

Why is the net cosmological constant so small?
The cosmological constant is connected with the symmetries of the metric $g_{\mu \nu}$. 
With a finite cosmological constant Einstein's equations of gravitation have no vacuum solution where $g_{\mu \nu}$ is the constant Minkowski metric.
That is, 
global spacetime translational invariance of the vacuum is broken by a finite value of $\Lambda$
\cite{Weinberg:1988cp}.
The reason is that a non-zero cosmological constant acts as a gravitational source which generates a dynamical spacetime  
 with accelerating expansion of the Universe for positive $\Lambda$.
Suppose that the vacuum including condensates with finite vacuum expectation values is spacetime translational invariant and that  
flat spacetime is consistent at mass  dimension four, 
just as suggested by the success of the SM.
With the SM as an effective theory emerging in the infrared, 
the low-energy global symmetries including spacetime translation invariance can be broken through additional higher dimensional terms 
suppressed by powers of the large scale of emergence $M$.
QCD and electroweak interactions are characterised by the scales  
$\Lambda_{\rm QCD} \approx 300$ MeV and 
$\Lambda_{\rm ew} \approx 246$ GeV. 
These scales might then enter the cosmological constant
with the scale of the leading term suppressed by the factor  
$\Lambda_{\rm ew}/M$ 
 -- see Refs.~\cite{Bass:2020nrg,Bass:2020egf} and the early work
\cite{Bjorken:2001pe,Bjorken:2001yv}.
This scenario, if manifest in nature, 
would explain why the cosmological constant scale $\mu_{\rm vac} = 0.002$ eV 
is similar to what we expect for light neutrino masses  
\cite{Altarelli:2004cp} assuming that neutrinos are Majorana with masses given by the Weinberg operator~\cite{Weinberg:1979sa}.
One finds
\begin{equation}
 \mu_{\rm vac} \sim 
 m_\nu \sim \Lambda_{\rm ew}^2/M   .
 \label{eq6}
\end{equation}
In this scenario, the cosmological constant would vanish at mass dimension four.
This is equivalent to a renormalisation condition $\rho_{\rm vac} =0$
at $D=4$ 
imposed by global space-time translational invariance of the vacuum, 
even in the presence of QCD and Higgs condensates, and keeping with the idea of emergent symmetry in the infrared.
Taking the value 
$\mu_{\rm vac} = 0.002 {\rm \ eV}$
from astrophysics together with $\Lambda_{\rm ew} \approx 246$ GeV gives the value $M \sim 10^{16}$ GeV.
With only the QCD, and electroweak scales in the picture, 
the largest term we can get is the $\Lambda_{\rm ew}^2/M$ factor 
with the DE scale entering at the same order in a low energy expansion as possible Majorana neutrino masses.
The large scale $M$ is again within the range where the SM Higgs self-coupling $\lambda$ might cross zero in the ultraviolet.
It is also similar to the value that is typically taken for the scale of primordial inflation \cite{Baumann:2008bn}. 
Might there be a connection?

Possible time dependence of DE is a subject of topical interest with hints from recent DESI collaboration measurements~\cite{DESI:2025zgx}.
Theoretical ideas include the vacuum expectation value of a possible 
time dependent extra scalar ``cosmon'' field
which interpolates between initial inflation and dark energy today~\cite{Wetterich:1987fm,Wetterich:1994bg,Peebles:1987ek,Peebles:2002gy} 
as well as running vacuum models 
\cite{SolaPeracaula:2022hpd}
and quantum breaking arguments~\cite{Dvali:2017eba}.
Within the emergence scenario here, 
possible time dependence 
might occur in the scale of emergence or ultraviolet completion $M$ 
and/or in the coefficient of the
$\Lambda_{\rm ew}^2/M$ term that appears in the 
low-energy expansion. 
This time dependence would reflect  relaxation of 
$\rho_{\rm vac}$
(and the SM) 
as one gets time-wise further away from the early Universe phase transition that produced it.
Condensed matter analogies for time dependent DE are described in Refs.~\cite{Volovik:2004gi,Volovik:2006bh,Volovik:2023faj}.

With emergence, the small size of the cosmological constant scale compared to the large scales of SM particle physics can be understood as follows.
From Eq.~(\ref{eqn15}),
the net
$\rho_{\rm vac}$ receives contributions from ZPEs, potentials in the vacuum and the net 
``bare gravitational'' term $\rho_\Lambda$.
Only the net cosmological constant
(plus any time dependence)
is the observable with the different sub-contributions
intermediate steps in the calculation.
The net scale is set by global spacetime translational invariance of the vacuum plus breaking by SM related terms in higher dimensional operators.
The ZPE and potential terms become cancelled by the $\rho_\Lambda$ 
contribution to preserve global 
spacetime translational invariance of the vacuum in the low-energy system characterised by the SM \cite{Bass:2023ece}.
A similar effect occurs in condensed matter physics with the Gibbs-Duhem relation for quantum liquids. 
Here the zero-point energy from low temperature  quasiparticles is cancelled by the effect of macroscopic degrees of freedom above the characteristic energy for the quantum liquid effective theory, 
e.g., atoms  \cite{Volovik:2004gi}.
The quantum liquid vacuum 
as well as simple Ising systems in the absence of an external magnetic field
\cite{Bass:2014lja} satisfy a vacuum equation of state.
The cosmological constant does need not jump when we go through the QCD phase transition or crossover transition in the early Universe.
Any change in the ZPE and potential terms might be  compensated by changes in the gravitational term 
$\rho_\Lambda$, with the latter term  
interpreted as parametrising the effect of physics above the scale of emergence.

\subsection{The Higgs mass and scale hierarchies in the SM}

The Higgs boson's mass 
is very much less than
the Planck scale 
despite quantum corrections which naively act to push its mass towards the deep ultraviolet.
Under renormalisation, the Higgs boson's mass squared comes with a quadratically divergent counterterm 
which comes from the Higgs boson self-energy, viz. 
\begin{equation}
m_{{\rm H \ bare}}^2 
= m_{{\rm H \ ren}}^2 + \delta m_{\rm H}^2,
\label{eq:4a}
\end{equation}
where
\begin{eqnarray}
\delta m_{\rm H}^2 
&=&
\frac{K^2}{16 \pi^2}
\frac{6}{4} 
\biggl(
8 \lambda 
+ 3 g^2 + g'^2 
-8 y_t^2
\biggr) + ...
\nonumber \\
&=&
\frac{K^2}{16 \pi^2}
\frac{6}{v^2} 
\biggl(
m_{\rm H}^2 + m_{\rm Z}^2 + 2 m_{\rm W}^2 - 4 m_t^2
\biggr) + ...
\label{eq:4b}
\end{eqnarray}
relates the renormalised and bare Higgs boson masses
and we neglect small contributions from lighter mass quarks.
Renormalised quantities are those dressed
by interactions and measured in experiments.
The corresponding bare quantities are taken direct from the Lagrangian and correspond to their values 
defined at the ultraviolet cut-off for the theory.
In Eq.~(\ref{eq:4b})  
$K$ is an ultraviolet cut-off scale on the momentum integrals characterising the limit to which the Standard Model should work.
If $K$ is taken as a physical scale, e.g., the
Planck scale,
then why is the physical Higgs boson's mass so small compared to the cut-off?

This hierarchy or
naturalness puzzle 
has attracted much theoretical attention~\cite{Giudice:2008bi,Wells:2009kq}.
What stabilises the value of $m_{\rm H}$? 
One possibility is that the Higgs boson's mass is fine tuned, perhaps through some kind of environmental selection and perhaps in connection with the vacuum stability of the Standard Model. 
Alternatively,
the Standard Model
quantum correction to the Higgs boson's mass, 
which is dominated 
by the top quark contribution,
might be cancelled by any new particles that couple to the Higgs boson.
However, such particles have so far not been seen 
in the mass range of the LHC.
Also the different terms on the right hand side of Eq.~(\ref{eq:4b}) satisfy different RG behaviour so if they cancel at one scale, then they do not necessarily cancel at others.
Likewise, any composite structure to the Higgs boson would soften the ultraviolet divergences 
but there is no evidence for this in the present data.
Searches for extra 
particles and possible composite structure will continue in the next years with increased luminosity at the LHC. 
The hierarchy puzzle is 
linked to the ultraviolet behaviour of the theory.
If one instead calculates using dimensional regularisation and 
$\overline{\rm MS}$,
then 
there is no large scale to cut-off the momentum integrals and the divergence appears as a pole in the dimensional continuation.
Thus, the hierarchy problem becomes hidden in the details of the calculation.
Actually, in this case 
one finds that the quadratic divergence term is ``thrown away" in the dimensional regularisation procedure.
However, if we treat the SM as an effective theory, then masses not protected by some key symmetry property like gauge invariance or chiral symmetries are expected to be large, of order the characteristic energy for the effective theory. So we still have a naturalness puzzle \cite{Wells:2009kq}.
In our emergence scenario the Higgs mass is environmentally  constrained by the requirement of vacuum stability \cite{Bass:2023ece}.

In thinking about the Higgs mass hierarchy puzzle with electroweak symmetry breaking and the Higgs mass and vev values 
$m_{\rm H}, v \ll M_{\rm Pl}$, 
it is interesting 
to consider  
similar phenomena with the ferromagnetic 
phase transition in condensed matter physics. 
Below the phase transition the magnetisation is very small 
close to the phase transition when we approach the critical temperature $T_c$ from below, viz. 
when the reduced temperature
$(T-T_c)/T \to 0^-$. 
Whereas emergent gauge systems can be associated with topological like
phase transitions without a local order parameter, 
Higgs phenomena is associated with spontaneous symmetry breaking 
with a
local order parameter which is
defined with respect to a particular gauge choice~\cite{Kibble:2014gug} with all choices of gauge being equivalent.
Electroweak symmetry breaking might correspond to a Universe 
close to the phase transition and very near to the critical point \cite{Jegerlehner:2013cta}, as also hinted at with electroweak vacuum stability.

In Eq.~(\ref{eq:4b}) the counterterm 
is negative
$\delta m_{\rm H}^2 <0$ 
is we substitute the measured masses and couplings from the LHC.
In general, 
with running masses and couplings, 
this $\delta m_{\rm H}^2$ 
implies that the bare mass squared is negative for energy scales less than some huge scale in the ultraviolet. 
This corresponds to spontaneous symmetry breaking,  
perhaps always below the scale of emergence. 
An interesting question is whether the counterterm can 
change sign at some large scale with running couplings, 
so called Veltman crossing \cite{Veltman:1980mj}.
Any sign flip in $\delta m_{\rm H}^2$ would also flip the sign of the bare mass squared term and thus give rise to a symmetric phase for this term
which would then persist to even higher energies
if we can extrapolate the perturbative SM up to these scales. 
The energy scale where this might happen is calculation dependent
\cite{Jegerlehner:2013cta,Bass:2020nrg,Degrassi:2012ry,Masina:2013wja,Hamada:2012bp}
with numbers quoted ranging between about $10^{15}$ GeV
\cite{Jegerlehner:2013cta} up to energies 
much above the Planck scale \cite{Degrassi:2012ry}. 
(Extra numerical corrections are also possible from the effect of higher dimensional terms at large scales.)
In the early Universe at
finite temperature $T$ there is an interesting issue associated with the Higgs potential \cite{Jegerlehner:2013cta}.
This potential generalises to
\begin{equation}
    V(\Phi, T) = \frac{1}{2} (g_T T^2 - \mu^2) \Phi^\dagger \Phi + \frac{1}{4} \lambda (\Phi^\dagger \Phi)^2+ ...
\end{equation}
Here
$\mu^2 = -m^2$ in Eq.~(\ref{eqn6}) and 
$g_T = \frac{1}{4v^2} (2 m_{\rm W}^2 + m_{\rm Z}^2 + 2 m_t^2 + \frac{1}{2} m_{\rm H}^2)
= 
\frac{1}{4} ( \frac{3}{4} g^2 + \frac{1}{4}g'^2 + y_t^2 + \lambda )
$.
Restoration of electroweak symmetry behaves differently depending on whether 
$\mu^2$ 
should be taken as the renormalised or bare mass squared here.
A sign flip in the term
$(g_T T^2 - \mu^2)$
is taken as a signaling  restoration of electroweak symmetry. 
The difference between taking $\mu^2$ as the renormalised or bare values corresponds to 
either a crossover at 
scales of a few hundred GeV or a first order phase transition at $10^{16}$ GeV.
The latter might be seen as stochastic background in future 
high frequency, 
few GHz,  gravitational waves measurements. 

\section{Parallels between particle physics and condensed matter}
\label{sec:TCM}

We have highlighted emergence as a possible underlying principle behind the success of the SM in our present experiments.
Emergent gauge symmetries are well known in condensed matter physics.
Analogies with condensed matter systems have previously inspired new thinking in particle physics.
There are analogue phenomena between the BEH mechanism in particle physics and massive 
``plasmons" in superconductors.
Various QCD confinement ideas and dynamical chiral symmetry breaking in 
low-energy QCD have parallels in condensed matter systems.
We next briefly explain these connections and then discuss emergent gauge systems in condensed matter physics which might motivate new thinking about the deeper structure of matter.

First, traditional superconductors are described by BCS theory \cite{Schrieffer:1964zz}.
In normal superconductors 
electron Cooper pairs are formed via phonon electron interactions \cite{Bardeen:1957mv}, with the phonons linked to lattice vibrations 
and not governed by gauge principle dynamics.
These BCS Cooper pairs can condense in a Bose-Einstein condensate 
with associated spontaneous symmetry breaking \cite{Nambu:1960tm}.
The Cooper pairs lead to  screening currents 
expelling 
the magnetic field 
in superconductors
(the Meissner effect).
The electron quasiparticles develop a mass gap from interaction with the condensate. A parallel in low energy QCD is 
the Nambu-Jona-Lasino model \cite{Nambu:1961tp,Nambu:1961fr} 
with massive constituent quark quasiparticles forming from coupling to the chiral quark condensate.
When photons propagate in superconductors, 
they become 
massive ``plasmons" with the photon as a wave on a sea of BCS Cooper pairs \cite{Anderson:1963pc}.
One finds a 
non-relativistic precursor to the BEH mechanism in particle physics.

Superconductors come in types I and II. 
In type I the magnetic field is expelled. 
In type II superconductors one finds partial entry of the external field within flux tubes. The material is superconducting outside the flux tubes but not inside. 
The type II superconductor acts like filaments of magnetic flux drilled in a type I material with a vortex of screening current surrounding each filament of magnetic flux. 
A gedanken isolated magnetic 
monopole would have infinite energy in a superconductor because of the Meissner effect. Any monopole is a source of magnetic flux, but this flux becomes compressed into a flux tube, with an energy proportional to its length. The magnetic flux out of any closed surface is quantised. A monopole anti-monopole pair would be joined by a quantised magnetic flux filament.
Type II superconductors have inspired ideas about QCD confinement.
The key idea involves so-called dual superconductors and condensation of colour magnetic monopoles plus formation of colour flux tubes. 
With a QCD colour dual superconductor the colour ${\bf E}$ and ${\bf B}$  fields are interchanged. 
One finds topological monopole like gluon configurations which condense in the vacuum. 
This leads to the formation of colour electric flux tubes -- QCD or hadronic strings 
linking colour carrying quarks with an accompanying string tension and connection to Regge trajectories.
In this ’t Hooft-Mandelstam confinement mechanism 
the  quarks are confined due
to these flux tubes
linking the quarks and 
preventing their separation, with QCD colour fields expelled from the vacuum much like magnetic fields are expelled from a 
superconductor \cite{tHooft:1977nqb,tHooft:1979rtg,Mandelstam:1974pi}.

Besides dual superconductors, there are also ideas exploring parallels between Andreev reflection 
and Kondo effect with confinement \cite{Gribov:1986vp,Sadzikowski:2002sk}. 
Andreev reflection \cite{Andreev:1964,degennes:1963,stjames:1964} 
concerns the interface of a normal metal and a  superconductor.
When an electron from the metal is incident on the junction surface, it is reflected back into the metal as a hole with a two-electron  Cooper pair system transmitted into the superconductor. 
This resembles the production of a quark-antiquark pair in $e^+ e^-$ annihilation with the region between the quarks playing the role of the metal and the external region or vacuum playing the role of the superconductor with propagating mesons. When the QCD running coupling reaches a critical value the quark ``transforms" into a meson and a reflected quark. There is an important difference in that Andreev reflection takes place just in a small momentum interval near the Fermi surface whereas quark confinement happens at all momentum values with small virtualities. 
A second confinement analogy involves the Kondo effect \cite{Kondo:1964nea}. When the QCD coupling reaches a critical value, it then drops to zero with the appearance of colourless hadron bound states. With the Kondo effect magnetic impurities in a metal interact with surrounding electrons to completely screen the spin of the impurity forming a non-magnetic singlet state.

Condensed matter ideas have also inspired thinking about the QCD phase diagram at finite densities and temperatures, 
see Ref.~\cite{Rajagopal:2000wf}.

The above condensed matter systems involve 
Landau-Ginzburg type phase transitions characterised 
by local order parameters.
Landau-Ginzburg type phase transitions in particle physics include 
dynamical chiral symmetry breaking associated the quark condensate in the QCD  vacuum 
and electroweak symmetry breaking  
(though there the order parameter is not gauge invariant but all choices of gauge are physically equivalent). 
In condensed matter physics  there are also phase transitions which come with no local order parameter, so called
topological phase transitions, 
and long range quantum entanglement. 
These phase transitions give rise to new topological phases of matter. 
The ground state can exhibit multiple degeneracy with the degenerate 
substates being related by emergent gauge transformations. 
Collective gauge fields beyond the more fundamental photons of QED 
can ``emerge'' from the quantum structure of the many-body ground state.  
The prototype emergent gauge system in condensed matter physics involves  
the low energy limit of the Fermi-Hubbard model of strongly correlated electron systems developed by 
Anderson and collaborators
\cite{Baskaran:1987my,Affleck:1988zz}.
This model is important in
many ideas about high temperature superconductors,
which require a new mechanism beyond traditional BCS theory, 
and quantum spin liquids
\cite{Powell:2020osu,Sachdev:2015slk}.
Another interesting emergent gauge system involves the A-phase of superfluid $^3$He.
In the vicinity of a Fermi point 
the quasiparticles
include gapless 
chiral fermions 
coupled to emergent U(1) and SU(2) gauge fields 
\cite{Volovik:2003fe,Volovik:2008dd}.
One also finds emergent  Lorentz invariance with limiting quasiparticle velocities 
and an emergent gravity like force.
Another emergent gauge system involves 
string nets where one finds a unification of gauge symmetries and spin-statistics~\cite{Wen:2007joe,Levin:2004js}.
Other examples of emergent gauge symmetries include 
spin ice in magnetic systems~\cite{Rehn:2016eqc} and 
the quantum Hall effect \cite{Tong:2016kpv} 
plus various more simple quantum systems~\cite{Wilczek:1984dh}.
Emergent gauge fields play an essential role in quantum simulations of quantum field theories~\cite{Banerjee:2012pg,Zohar:2015hwa}.
We next illustrate key condensed matter emergent gauge systems to explain how the emergent symmetries appear.
With a possible emergent particle physics SM, there is a fundamental difference
with emergent condensed matter systems in that there we know the degrees of freedom above and below the phase transition whereas with an emergent particle physics we measure only the low-energy phase in our experiments.

\section{Emergent gauge symmetries in condensed matter systems}
\label{sec:GSCM}

\subsection{The Fermi–Hubbard model and its low energy limit}

In condensed matter physics the prototype emergent gauge system is the Fermi-Hubbard model of strongly correlated electrons in a two dimensional atomic lattice at half-filling.  
In the low-energy limit the system exhibits SU(2) and U(1) emergent gauge symmetries and spin-charge separation~\cite{Baskaran:1987my,Affleck:1988zz}. 
The emergent SU(2) couples to the spin of the electrons which then becomes dynamical to internal observers.
This model is a starting point for many discussions of high temperature  superconductors. 
Usual superconductivity occurs up to about 4K whereas high temperature superconductors refer to new phenomena discovered in ceramic cuprates at temperatures about 77K.

The Fermi–Hubbard model describes electron behaviour in an atomic lattice.
One considers a lattice of 
atoms supporting only a single atomic state, which can hold up to two electrons with opposite spins.
The electrons interact with the potential of a static lattice of ions.
One neglects any motion of the ion lattice, being only interested
in interactions of electrons and not in dynamical lattice effects, such as phonons. The electrons interact via Coulomb repulsion.
Further, one assumes that all except the lowest band have very high energies and are, thus, energetically unavailable.
Also, that the remaining band has rotational symmetry.
The electron hopping matrix, 
which describes electron motion from
one lattice site to another,
depends just on the distance between lattice sites.
Finally, one restricts the model to nearest neighbour interactions
(with underlying matrix elements decreasing fast with increasing distance).
Any additional fluctuations of the atomic lattice sites corresponds to extra bosonic phonon excitations beyond the emergent gauged system independent of whether the lattice ions are fermions or bosons.

The Fermi-Hubbard model Hamiltonian then has two terms:
a hopping term between nearest neighbour sites 
with coupling strength $t$,
plus an ``on-site'' Fermi–Hubbard repulsion term $U$,
\begin{equation}
    {\cal H} = - t \sum_{(ij)\sigma} c_{i \sigma}^\dagger c_{j \sigma} + U \sum_i 
    c_{i \uparrow}^\dagger c_{i \uparrow} c_{i \downarrow}^\dagger c_{i \downarrow} .
\label{eq:6.1}
\end{equation}
Here a 
square lattice is assumed where $ij$ are nearest neighbour bonds.
$c^\dagger_{i \sigma}$ and $c_{i \sigma}$ 
are the creation and annihilation Fock operators for electrons 
with spin $\frac{1}{2}$
on site $i$.
The first term prefers 
non-localisation whereas the second prefers localisation
with just one electron on each lattice site.
Extra ``doping'' terms can be included by adding a chemical potential.
In the low-energy Mott limit
$U \gg t$ 
the Fermi-Hubbard system behaves as an insulator.
(With extra doping terms described by adding a chemical potential, it becomes a model for describing high temperature superconductors.)
Treating the hopping term as a perturbation and
keeping the leading term evaluated using Rayleigh-Schr\"odinger perturbation theory, 
the Fermi-Hubbard model
reduces to the Heisenberg magnet Hamiltonian.
For the half filled system one finds
\begin{equation}
    {\cal H}_{\rm eff} =
    J \ \sum_{i,j} 
    (c_{i \alpha} ^{\dagger} \sigma_{\alpha \beta} c_{i \beta})
    .
    (c_{j \alpha} ^{\dagger} \sigma_{\alpha \beta} c_{j \beta})
\label{eq:6.2}
\end{equation}
where $J=4t^2/U$,
the $\sigma$
denote SU(2) Pauli matrices, 
and one has the constraint
$c^{\dagger}_{j \alpha} c_{j  \alpha}=1$.

The electron quasiparticles here exhibit 
spin-charge separation.
Emergent U(1) and SU(2) 
gauge symmetries appear, with the latter coupled 
to the spin degrees of freedom of the electrons.
One finds accompanying $J=1$ gauge bosons among the quasiparticles in the strongly coupled electron system.
Formally, spin-charge separation in 
the strongly correlated electron system can 
be described using
a slave-particle
representation 
using auxiliary fermion and boson operators.
One writes
the $c$-electron Fock operators as
a combination of 
``spinon''
$f$-electrons carrying spin and no electric charge, 
and spinless ``holons''
which carry the electric charge, that is, with spin-charge separation.
In the low-energy Fermi-Hubbard model
with half filling
the spin operators appearing in the product in Eq.~(\ref{eq:6.2})
are chargeless and
the low energy system can be written just in terms of 
the $f$-electrons,
$c \mapsto f$ in Eq.~(\ref{eq:6.2}).
The Hamiltonian 
(\ref{eq:6.2})
has the important local gauge symmetry 
$f_{j \sigma}^{\dagger}
\to 
e^{i \theta_j}
f_{j \sigma}^{\dagger}$.
The model system Eq.~(\ref{eq:6.2}) also exhibits a local SU(2) gauge symmetry.
To see this, first consider the electron operators
$(f_1, f_2)$
and
$(f_2^\dagger, -f_1^\dagger)$
which transform as SU(2) spin doublets.
These are combined to form the matrix
\begin{equation}
\Psi =
\left(\begin{array}{cc}
f_1 & f_2
\\
f_2^\dagger &
- f_1^\dagger
\end{array}\right) 
\label{eq:6.4}
\end{equation}
which transforms under global SU(2),
$
\Psi \to \Psi g$.
One can define a second local SU(2)
symmetry by
$\Psi \to h \Psi$.
Here $g$ and $h$ denote 
SU(2) rotations,
viz.
$e^{i \vec{\sigma}.\vec{\omega}/2}$
where 
$\vec{\sigma}$
denotes the SU(2) Pauli matrices and
$\vec{\omega}$ 
is spacetime independent
for $g$ and spacetime dependent for $h$.
Spin operators for global SU(2)
can be written
${\rm S} = \frac{1}{2} \Psi^\dagger \Psi \sigma^{\rm T}$
where $\sigma^{\rm T}$ is the transpose of $\sigma$.
Since
$
\Psi^\dagger \to g^\dagger \Psi^\dagger h^\dagger
$
it follows that the 
spin operators are invariant under local SU(2).
That is, the Heisenberg interaction
is invariant under local SU(2) gauge transformations
with $h$ denoting an element of the gauge group.
The Hamiltonian in Eq.~(\ref{eq:6.2})
can be written in terms of the spin operators 
as
\begin{equation}
    {\cal H}_{\rm eff} 
    =
     J/4 \ \sum_{i,j} 
    ({\rm tr} \ \Psi_i^\dagger \Psi_i \sigma^{\rm T})
    .
     ({\rm tr} \ \Psi_j^\dagger \Psi_j \sigma^{\rm T}) .
\label{eq:6.5}
\end{equation}
The local gauge symmetry within the Heisenberg model acts trivially on the spin operators but becomes interesting within the 
large $U$ limit of the Fermi-Hubbard model with electron operators at half filling. 
This is a consequence of the redundancy of parametrising
spin operators by
electron operators.
Note that it is the ``spinon'' $f$-electrons 
that feel the emergent 
SU(2) gauge symmetry here
rather than it being a 
property of the charged
$c$-electrons of QED
which appear in the more general Fermi–Hubbard Hamiltonian Eq.~(\ref{eq:6.1}).
The emergent gauge symmetry seen here comes with an energy barrier.
Local SU(2) gauge invariance is valid up to below the 
Mott-Hubbard energy gap.
For large but finite $U$ there is an approximate gauge symmetry in the sense that it is only broken in the sector of the Hilbert space containing high energy states with energies of order $U$.

This physics leads to the 
dynamics of the resonating valence bond (RVB) model \cite{Anderson:1987gf}.
The Hamiltonian 
Eq.~(\ref{eq:6.2}) 
can also be expressed in the form \cite{Baskaran:1987my}
\begin{equation}
    {\cal H}_{\rm eff} =
    J \sum_{ij} 
    b_{ij}^{\dagger} b_{ij}
\label{eq:6.3}
\end{equation}
where
$b_{ij}^{\dagger}
=
(1/\sqrt{2})
(f_{i \uparrow}^{\dagger}
f_{j \downarrow}^{\dagger}
-
f_{i \downarrow}^{\dagger}
f_{j \uparrow}^{\dagger}
)
$
are bosonic 
single (two electron) creation and annihilation operators involving the ``spinon'' quasiparticle excitations.
These operators are important in the RVB theory of high temperature superconductivity in cuprates
\cite{Baskaran:ssc,anderson:prl87}.
The two ``spinon'' electron excitations 
behave as Cooper pairs and
can form a Bose-Einstein condensate.
The Fermi–Hubbard system exhibits (long range) entanglement and quantum correlations in its ground state 
in the $U \gg t$ limit 
\cite{Powell:2020osu}.
RVB states as a superposition of all nearest neighbour bond configurations can exhibit topological order characterised by long-range quantum entanglement
(independent of microscopic details of the Hamiltonian). 
One finds a new spin-one emergent gauge field as a collective effect in the strongly coupled electron system.
Including this gives the net Lagrangian
\begin{equation}
    {\cal L} = \frac{1}{2} \sum_{j} {\rm tr} \
    \Psi_{j}^\dagger \biggl( i \frac{\partial}{\partial t} + B_{j} \biggr) \Psi_{j} - {\cal H}_{\rm eff} 
\label{eq:6.6}
\end{equation}
with the gauge field
$
B = \frac{1}{2} \sigma.{\bf B}$
transforming as
$B \to h ( B + i (\partial/\partial t) ) h^{\dagger}$
under the local SU(2) gauge transformations associated with $h$.
The three components of ${\bf B}$ act as 
Lagrange multipliers and guarantee 
the half filled system with 
constraint of
one particle per site
\cite{Affleck:1988zz}.
Additional doping allows the possibility of d-wave pairing 
which can lead to new processes of superconductivity beyond traditional BCS theory. 
We refer to Ref.~\cite{Sachdev:2018ddg} 
for detailed discussion of the phase diagram of the Fermi–Hubbard model including confinement and Higgs phases,
as well as application to 
high temperature superconductors.

\subsection{Superfluid $^3$He-A}

In the A-phase of superfluid $^3$He 
the quasiparticles in the vicinity of a Fermi point are chiral fermions interacting with emergent U(1) and SU(2) gauge bosons 
\cite{Volovik:2003fe,Volovik:2008dd,Volovik:2010vx}.
Superfluidity works differently for fermionic $^3$He and bosonic $^4$He.
With $^4$He a superfluid forms at around 2K and involves s-wave Cooper pairs.
Superfluid $^3$He involves 
p-wave Cooper pairs and superfluidity sets in around 2 mK, much below the 2K with $^4$He~\cite{Vollhardt:2000ei}.
In the absence of an external magnetic fields
there are two phases called 
the A and B phases.
In the B-phase the Cooper pairs have $J=0$ and the quasiparticle energy gap is isotropic.
In the A-phase the Cooper pairs have just $S_z =\pm 1$ states with the $S_z=0$ state absent.
Here the energy gap depends on 
the angle between the quasiparticle momentum and the orbital angular momentum of the Cooper-pairs, which is the same for all pairs.
Especially interesting is that
one find Fermi points, singular points in momentum space where the gap vanishes.
Close to these Fermi points the A-phase exhibits emergent Weyl chiral fermions together with emergent U(1) and SU(2) gauge symmetries linked to the topology in momentum space and an emergent metric with limiting quasiparticle velocities 
plus spin-two effective 
``gravitons''.  
That is, the structure looks like the SM in a symmetric phase with vanishing Higgs condensate.
The A-phase forms at 2.6 mK and 21 bars of pressure. 
The transition between the 
superfluid A and B phases of $^3$He 
is first order whereas the transition to superfluidity itself is second order~\cite{Tian:2023,QUEST-DMC:2024crp}.

The emergent quasiparticles 
and accompanying gauge symmetries are associated with 
the Fermi point and 
the freedom in choosing  the position of this Fermi point on the former Fermi surface.
Consider the fermion 
quasiparticle propagator
${\cal G}$ 
in the vicinity of the Fermi point $p=p^{(0)}$, viz.
\begin{equation}
 {\cal G}^{-1} (p_0)
 =
 i \omega - H ({\rm p})
 = 
 e_i^{\ k} 
   \Gamma^i  
   (p_k - p^{(0)}_k)    
 + \ {\rm higher \ order \ terms}. 
\label{eq:6.10}
\end{equation}
Here $\Gamma_i = (1, \sigma_i)$ and
the matrix $e^k_i$ is the analogue of the dreibein with 
$g^{ik}$ 
playing the role of an effective dynamical metric in which fermions move along geodesic lines. 
The Green's
function has a singularity at the Fermi point where 
the fermion quasiparticles are gapless with energies
$E = c {\bf \sigma .p}$ where $c$ is the limiting velocity. 
``Higher order terms'' represent additional contributions slightly away from the Fermi point. 
At the Fermi points one finds that the quasiparticle propagators 
are characterised 
by the
topological invariant quantity
\begin{equation}
    N_3 = {\rm tr} {\cal N},
    \ \ \
    {\cal N} = \frac{1}{24 \pi^2} \epsilon_{\mu \nu \lambda \gamma}
    \int_S
    {\rm d}\sigma^{\gamma} \ {\cal G \partial_{P_\mu} G}^{-1} {\cal G \partial_{P_\nu} G}^{-1} {\cal G \partial_{P_\lambda} G}^{-1} .
\label{eq:6.11}
\end{equation}
which is quantised in integer units. 
Here $S$ denotes a three dimensional surface around the isolated Fermi point and one takes the trace over relevant spin indices.
The physics is invariant 
under changes in the position of the Fermi point and this leads to an emergent gauge symmetry in the low energy system. 
The case  
$N_3 = \pm 1$ corresponds to 
U(1) and $N_3 = \pm 2$ to SU(2).
The Fermi point behaves as a hedgehog in momentum space with plus/minus signs of 
$N_3$
corresponding to the spins
pointing out/in.
Freedom to choose the position of the Fermi point corresponds to an emergent gauge symmetry. 
The gauge symmetries correspond to the freedom 
(degeneracy) in choosing the position of the Fermi point on the former Fermi surface.
One finds emergent local gauge interactions with, for SU(2), the spin of the $^3$He quasiparticles becoming dynamical to internal observers. 
The quasiparticles in $^3$He-A, both fermions and gauge bosons, 
each come with a common limiting velocity behaving like a relativistic quantum field theory like what happens with Lorentz invariance in the SM (for details see Sect. 9.3.2 of Ref.~\cite{Volovik:2003fe}). 

\subsection{String-nets}

In the context of particle physics,
another interesting system is that of
3+1 dimensional string-nets which
involve qubit chains in a lattice environment 
\cite{Levin:2004js,
Wen:2007joe}.
These chains can condense with excitations 
providing a model for 
electrons and photons
as emergent degrees of freedom linked to details of long range quantum entanglement.
Beyond condensation of bosonic qubits, 
one finds string excitations of connected bosons. These strings can condense as well as the individual bosons.
Excitations above this string condensate are either gauged bosons or fermions (as the ends of the strings).
In this system isolated fermions without gauge charges are not allowed. 
The emergent gauge symmetry is linked to 
spin-statistics properties of the different excitations.
The model can be extended from U(1) ``photons" to 
SU(N) gauge symmetries with ``quarks and gluons". 
The extra step of
obtaining chiral fermions remains an open puzzle in this approach, perhaps connected to the lattice input.

\section{Conclusions and open questions}
\label{sec:Conc}

The Standard Model and General Relativity are working very well in our present experiments.
High-energy particle physics is driven by the quest to understand Nature at a deeper level and 
the dynamics of the very early Universe.
How high in energy and precision will the present theories continue to work before we encounter new physics?
One key question is the origin of the gauge symmetries that drive the SM dynamics.

Here we have explored the paradigm of an emergent SM, ``born" below some phase transition deep in the ultraviolet.
The high-energy extrapolation of the SM is driven by the renormalisation group evolution of SM parameters: the three gauge couplings, the Higgs self-coupling and the particle masses that come from the Higgs sector.
The vacuum comes out within a few standard deviations of being stable up to the Planck scale and very close to the border of stable and metastable, which may hint at possible new critical  phenomena in the ultraviolet.
With an emergent SM, the physics behaves as an effective theory with 
characteristic energy about $10^{16}$ GeV.
This scale is suggested by the values of light neutrino masses deduced from  oscillation experiments and by the size of the dark energy scale extracted from astrophysics.
It is also close to what is commonly taken as the scale of primordial inflation.

In the emergence scenario the dark energy scale and Majorana neutrino masses enter at the same order in a low-energy expansion.
The Higgs mass is 
environmentally determined in connection with vacuum stability.
Time dependent dark energy might correspond to relaxation of the SM away from the initial phase transition that produced it.
There are also interesting possibilities for dark matter. 
Some new, non-luminous,  matter is required by astrophysics to comprise 84\% of the matter budget of the Universe \cite{Baudis:2017avj,Bertone:2018krk}. 
Axions, if present, enter at $D=5$ and make an interesting particle candidate. 
Considering analogies with condensed matter systems,
one might also consider 
possible parallels between  DM and non-gauged phonons in strongly correlated electron systems and excitations of the system above the scale of emergence.
Beyond the strongly correlated electron interactions, 
phonon vibrations of the atomic lattice 
exhibit bosonic statistics independent of the fermionic or bosonic nature of the atoms on the lattice sites. 
If the Fermi-Hubbard model provides a prototype for thinking about emergent gauge symmetries in particle physics, then the extra phonon excitations might provide a useful analogy for thinking about dark matter.
This conjecture needs detailed investigation to see whether it might have a chance to work.
Further condensed matter inspired ideas for DM are discussed in Refs.~\cite{Jegerlehner:2023sfw,Klinkhamer:2016zzh}.

How can we test these ideas and also explore 
possible new physics in the deep ultraviolet?
At collider energies we would like a precise measurement of the Higgs 
self-coupling at TeV scale energies to check that the SM is really working here.
Precision measurements of SM parameters, especially the top and Higgs masses and the QCD coupling, will further constrain our understanding of electroweak vacuum stability through radiative corrections.
With emergence we expect small multiplets meaning that SUSY, two Higgs doublet models... would be disfavoured.
New physics would enter in higher dimensional operators and at very high energies, about $10^{16}$ GeV, 
close to the scale of emergence.
Experiments with Majorana neutrinos
(assuming that neutrinos are Majorana as expected here) including CP phases might be sensitive to physics at these high scales.
Perhaps there are tiny effects with baryon number (proton decays) and Lorentz violation waiting to be found.
The interface of future particle physics and gravitational waves measurements offers new possibilities of investigation and discovery \cite{Bass:2026wzo}.
Measurements of high frequency gravitational waves with frequency about 1 GHz would be sensitive to 
possible phase transitions at these scales 
if these are first order and with a signal to be seen as a stochastic gravitational wave background~\cite{Domcke:2024soc,Aggarwal:2025noe}.
There are no known astrophysical sources at these frequencies. 
As an example of a condensed matter system with emergent gauge symmetries, 
the transition between the 
superfluid A and B phases of $^3$He 
(the A-phase with Fermi points and emergent U(1) and SU(2) gauge symmetries)
is first order whereas the transition to superfluidity itself is second order~\cite{Tian:2023,QUEST-DMC:2024crp}.
In particle physics 
if radiative corrections to the Higgs self-energy were to cross zero and change sign at a scale below the Planck scale
(a proposal which is calculation dependent and sensitive to the details of radiative corrections~\cite{Jegerlehner:2013cta,Bass:2020nrg,Degrassi:2012ry,Masina:2013wja,Hamada:2012bp}) 
this would result in a first order phase transition~\cite{Jegerlehner:2018zxm}.
Similar physics could be generated with the 
temperature dependent version of the Higgs potential 
if we consider the sign flip associated with the squared bare mass parameter at some very high temperature in the early Universe \cite{Jegerlehner:2013cta}.
Another key observable involves the tensor-to-scalar ratio and 
B-modes in CMB polarisation which are believed to be induced by gravitational waves propagating in the inflationary period~\cite{Baumann:2008bn,Komatsu:2022nvu}.
If the SM (and perhaps General Relativity like the situation with $^3$He-A) 
might be emergent at a scale $\sim 10^{16}$ GeV, 
then the degrees of freedom above the scale of emergence and acting in the inflationary period  might be quite different. 
There are interesting challenges for experiments and theory.
If the emergence scenario really describes Nature,
then there are profound connections between the infrared and ultraviolet  waiting to be explored.

\section*{Acknowledgements}

I thank Michal Praszalowicz for the invitation to this stimulating meeting in Zakopane and 
Fred Jegerlehner, Janina Krzysiak and Stefan Pokorski for interesting discussions on the physics topics discussed in these lectures.

\end{document}